# A Context Aware Framework for IoT Based Healthcare Monitoring Systems

*Yousef Abuseta*
Computer Science Department
Faculty of Science, Zintan University
Zintan, Libya
Yousef.m.abusetta@gmail.com

*Abstract*—This paper introduces an investigation of the healthcare monitoring systems and their provisioning in the IoT platform. The different roles that exist in healthcare systems are specified and modeled here. This paper also attempts to introduce and propose a generic framework for the design and development of context aware healthcare monitoring systems in the IoT platform. In such a framework, the fundamental components of the healthcare monitoring systems are identified and modelled as well as the relationship between these components. The paper also stresses on the crucial role played by the AI field in addressing resilient context aware healthcare monitoring systems. Architecturally, this framework is based on a distributed layered architecture where the different components are deployed over the physical layer, fog platform and the cloud platform

I. INTRODUCTION

With the increasing number of chronic disease cases recorded and reported all over the world, there is an urgent need for the enhancement of the functioning and service provisioning of healthcare monitoring systems [1]. In [2] [3], it has been reported that more than 400 million people in the world are connected with diabetes and cardiovascular diseases which can cause serious complications such as heart failure, other irregular heart rhythms, heart attack, stroke, and kidney failure.

The continuous monitoring of patients is crucial not only for recovering and curing from illness conditions but also for anticipating and predicting the possible occurrence of a particular disease. Being proactive and anticipating health problem conditions helps in discovering these problems at an earlier stage which often increases the chance of preventing or at lease decreases the damage of these health problems.

The remarkable advancement in communication, networking and sensor technologies along with the storage, big data analytics and computation capabilities offered by the cloud platform in the IoT paradigm have provided a promising infrastructure for the healthcare monitoring systems to accomplish their fundamental tasks in a robust manner. Also, the emergence and availability of many advanced technologies such as wireless sensor network, wireless body area network, wearable and implanted sensor, has enabled the IoT paradigm to play a crucial role in solving existing problems related to healthcare monitoring systems. However, due to the long distance between the physical layer (where devices of the healthcare monitoring systems are operating) and the cloud platform (where most medical data analysis takes place), the latter alone falls short of providing the appropriate support for the real time acting needed by healthcare monitoring systems in particular and most of IoT based applications in general. To address this issue, some of the computation, storage and analysis capabilities are provided at the Fog computing platform. The main idea here is to allow devices to communicate directly with each other without the need to send data along the way to the cloud; this enables making crucial decisions in real time and also shields IoT medical applications from transmitting massive amount of patient data to the cloud which results in optimising the network bandwidth utilization.

It is also crucial for healthcare systems to provide their services in a context aware manner. Context awareness can help healthcare systems avoid taking unnecessary actions that might be resource intensive. It also guarantees, to a large extent, that the right actions and services are delivered at the right time.

In this paper, an investigation of healthcare monitoring systems and their provisioning in the IoT platform is provided. This paper also highlights the potential benefits that IoT can bring to healthcare systems as well as the challenges and risks of applying the IoT paradigm. More importantly, this paper introduces and proposes a context aware generic framework for the design and development of healthcare monitoring system in the IoT platform. In such a framework, the fundamental components of the healthcare monitoring systems are identified and modelled as well as the relationship between them. This framework is based on a distributed layered architecture where the different components spread over the physical layer, fog platform and the cloud platform.

The rest of the paper is organized as follows. Section II reviews some background issues related to our proposed architecture. Section III introduces our proposed architecture for IoT based healthcare monitoring systems. In section IV, reviews some of the previous works that have been conducted so far. The paper is concluded in section V while some suggestions for further work are provided in section VI.

## II. BACKGROUND

### A. Healthcare Monitoring Systems

The main objectives of healthcare systems are to save human lives and improve the quality of people's life. It is been reported in [4] that the expected life average has increased by five years in the past two decades. However, the sharp increase in the in-hospital healthcare cost and the population aging issue have called for more efficient healthcare systems [5]. This leads to several research directions that involve many fields including the healthcare, data analytics, wireless communication, embedded systems, and information security. Implantable and Wearable Medical Devices (IWMDs) are envisioned as key components of modern healthcare systems. Such devices enable non-invasive prevention, early diagnosis, and continuous treatment of medical conditions [6].

In addition, the emergence and wide acceptance of the IoT paradigm as well as the introduction of computing and networking platforms such as the Cloud and Fog computing have enabled traditional healthcare systems to replace in-hospital medical systems with Internet connected IWMD based systems. The latter systems are operated in the IoT platform [7].

Healthcare Monitoring Systems are a result of the convergence of ubiquitous computing and communication technologies. These systems have been proposed to provide e-health services that match the context and real needs of patients. The main objective of all of the proposed solutions for these systems is to provide a smart environment where the health conditions of patients (also elderly and disabled people) are monitored and evaluated and then e-health services are provided in a timely manner [8].

In Healthcare Monitoring Systems, three categories of users can be identified. They include the monitored element, medical professionals and caregivers which are described as follows:

- *Monitored element*. Entities such as the patient, elderly and disabled belong to this category. These entities are the central components of Healthcare Monitoring Systems.
- *Medical personnel*. This category contains entities responsible for providing medical services to monitored people (patient, elderly and disabled). Such entities include doctors, consultants, nurses, etc.
- *Caregivers*. This category represents any entity that has an interest in the monitored person's conditions or current situation.

### B. IoT architecture for Healthcare Monitoring Systems

A high level architecture of healthcare monitoring systems deployed on the IoT platform is introduced in this section. This architecture is based on a previous work presented in [9]. The following layers constitute the proposed architecture:

- *The sensing layer*: This layer is located at the bottom of the architecture and contains the physical devices and sensors responsible for sensing and collecting the patient data that is of great importance to the Healthcare Monitoring System (HMS). For example, a temperature sensor is used to collect the temperature readings of a patient.

- *The Fog layer*: This layer sits between the sensing layer and the IoT service layer. It serves as a gateway that abstracts away the heterogeneity of the communication protocols that each device supports. In addition, this layer is employed to provide some of the computation, storage and analysis capabilities that are needed by Healthcare Monitoring Systems. This is referred to in [ 9] as the local manager since it is located close to the sensing layer that contains the devices responsible for providing the required services.

- *The IoT service layer:* it is located at the Cloud platform and contains the computation, storage and analysis capabilities that are needed for the processing of data produced at the sensing layer by the physical devices. At this layer, more powerful computation, storage and analytics services are offered which can be seen as an extension to the services provided in the Fog layer.

- *The application layer:* this layer is concerned with providing applications that enable end users to interact with the healthcare system for the purpose of querying a patient condition, receiving notifications of patient status and emergency situations.

### C. Context Aware Systems

The context concept has been defined by my researchers in the literature. In [10], the context is defined as follows: "*Any information that can be used to characterize the situation of entities (i.e., whether a person, place or object) that are considered relevant to the interaction between a user and an application, including the user and the application themselves*".

Since the management of all the context information is unpractical and difficult to accomplish, context aware systems should adopt an appropriate mechanism to choose the relevant subset of the context information of the application under consideration.

Different dimensions and perspectives have been employed to classify and model the context concept. In [11], the authors used the location, identity, time and environment dimensions to represent the context. A similar classification introduced in [12] where the authors select the identity, location, activity and time dimensions to model the context. Zimmermann et al. [13] classified context information into five categories: individuality context, time context, location context, activity context, and relations context. In [14], authors use location, identity of nearby people and objects, and changes to those objects to model the context concept.

The use of the context concept in computing has led to the emergence of the context aware systems. In [10]," *A system is context-aware if it uses context to provide relevant information and/or services to the user, where relevancy depends on the user's task"*. According to [15], a context aware system is a system that provides relevant information or takes proper actions on the basis of the user's current context and need. Another definition by [16], states that a computing system is regarded as context aware if it is able to monitor inputs from sensors and chooses the right context according to the user's needs and interests.

The design process of context aware systems faces a number of challenges which can be described as follows:
- The collection of contextual information with the use of sensors (e.g. calendar, light, battery charge, etc.)
- The modeling of context.
- The reasoning about the context to produce an adaptive behavior.

D. *Context awareness in Healthcare Monitoring Systems*

This section highlights the important aspects of the context awareness property in relation to the engineering and design of Healthcare Monitoring Systems. As pointed in [17] the context is crucial for healthcare monitoring for two reasons. Firstly, the evaluation of the health conditions of monitored patients relies, to a large extent, on the understanding of the contextual data gathered from a set of sensors. Therefore, any significant changes in vital signs data should be associated with a better understanding of the current patient's situation. For example, interpretation about the cardiac activity should be linked to the nature of the current activity being performed by the patient. Secondly, the context can make the process of monitoring the patient health conditions more efficient and optimal. As known in the IoT platform, the data belonging to the patient's context, such as the behavioural, physiological, and environmental data, are often obtained through a set of sensors. The continuous processing of this data requires computation power and tends to cause network problems such as high consumptions of bandwidth and overloaded servers. The optimal solution here is to base the healthcare monitoring systems on the collection and analysis of high relevant contextual information. Context information in Healthcare Monitoring Systems can be broadly classified into three fundamental types, namely the behavioural, physiological and environmental.

Context awareness assists in a more accurate diagnosis of the health conditions of the monitored patient. It can identify the behavior patterns and thus make more precise inferences about the situation of persons and their environment. Pichler et al. [18] pointed out that the three most important benefits of context awareness are the adaptation, personalization, and proactivity. What follows is a description of these three benefits:

- *Adaptation:* it is concerned with adjusting a service or information according to the current context of the user. A concrete example is when the system in question adapts its delivered data with respect to the network and device context such as the connection speed and display resolution.
- *Personalization:* it deals with adapting a system to different users, such that different users see the system differently at the same time. The Personalization process is based on each user's preferences, habits, skills, tasks, etc. The information or, more precisely, the level of details at which information is provided to the doctor, for example, in Healthcare Monitoring Systems is obviously different from that delivered to the patient or caregivers.
- *Proactivity*: it is concerned with delivering services to the user on the basis of predictions of future contexts. With respect to Healthcare Monitoring Systems, Proactivity plays a crucial role in delivering very beneficial and promising solutions. Examples and use cases are countless. For instance, being proactive and anticipating health problem conditions helps in discovering these problems at an earlier stage which often increases the chance of preventing or at lease decreases the damage of these health problems. Another scenario or use case is where proactivity helps in predicting disease-causing mutations that are caused by genetic changes in the genome, which are likely to have a molecular effect.

E. *AI role in Context Aware Systems Design*

This section introduces the important role of the Artificial Intelligence field in addressing the design and implementation of Healthcare Monitoring Systems. As discussed earlier in this paper, the importance of the context awareness to the engineering of Healthcare Monitoring Systems is evident. To further boost the positive effect of context aware systems, there is a fundamental need for more intelligent systems that can demonstrate the required quality to interpret the collected data accurately and act accordingly by providing the user (patient, elderly or disabled) with services relevant to the current situation or context. Concepts and applied methods from the Artificial Intelligence (AI) field are regarded as highly promising solutions. For instance, context aware systems need to take into account the environment that surround them and to carry out the right actions in response to changing contexts. The environment is one of the fundamental concepts of the AI field and is a core component of the agent system model. Another issue that is of high importance to the context aware systems is knowledge representation. User preferences or profiles, which represent knowledge artifacts, tend to be difficult to express using traditional bivalent logic (a statement is either true or false) and therefore non classical logic constructs such as fuzzy logics can help in representing contexts [19]. The reasoning aspect, one of the pillars of the AI paradigm, is also of high significance to the context aware systems. The reasoning is concerned with selecting the appropriate adaptation on the basis of the available contextual information. It involves examining some conditions and if fulfilled appropriate actions are taken. A common forms of reasoning are of a logical and rule based nature.

### III. PROPOSED FRAMEWORK FOR HEALTHCARE MONITORING SYSTEMS

The proposed framework for the design and development of healthcare monitoring systems is presented here. IoT Applications presented in this paper are built on some concepts and models discussed in the background section. The architecture is viewed as consisting of two fundamental layers: the managed element and managing element. The following subsections are dedicated to introduce the design and architecture of these two layers as well as any justifications and explanations about any design decisions made in this proposed architecture.

#### A. *Characteristics of Proposed Framework*

For the proposed framework to be effective and resilient, a number of issues and requirements must be addressed and provided which are listed below:
- *Usability*: it should be easy to use and implement by developers of healthcare monitoring systems.
- *Generality*: it should be generic enough to be used across a variety of healthcare monitoring systems.
- *Extensibility*: it is related to the framework ability to be extensible in order to accommodate new features and capabilities. For instance, the framework should be flexible enough to introduce a new physical entity as well as new protocols that support these entities.
- *Customization:* it is concerned with the ability of the framework to be customized and tailored for some specific systems or some organizations of feedback control loops (e.g. centralized or decentralized) .
- *Prediction support*: the framework should provide the required software components responsible for predicting any potential occurrence of health problems.
- *Context awareness support*: it is concerned with providing the necessary components responsible for augmenting healthcare systems with context aware capabilities.
- *Testability*: it concerns with the ability of the framework to be tested for some tasks and activities. Testing the process of monitoring a specific system property and taking the appropriate corrective actions is only one example.

#### B. *Conceptual View of Proposed Framework*

This section introduces the metamodel of the healthcare monitoring system which includes the fundamental concepts and entities as well as the relationships between them. The principal entity of this model is the *monitored element* which can play the role of the patient, elderly or disabled. Each monitored element is assigned one or more sensors to take readings of the physiological data (e.g. temperature, blood pressure, etc) of interest to the healthcare monitoring system in question as well as the environmental data (e.g. room temperature) and activity data (e.g. moving, laying in bed, time of activity, location of activity, etc). These three categories of data constitute the *context* of the *monitored element.* The *context history* component keeps track of the context data of the monitored element so the decision making is based on a series of observations and not only on the current situation presented by the current context data. This data is utilized by the *reasoning engine* component to run the predictor and extract some insights that are of great interest to the monitored element's health conditions or situations. The *insight* component represents the health condition or a worth considering situation of the *monitored element*, a fall incident for instance. As soon as an insight is discovered, it is passed to the *rule engine* component to compose the appropriate corrective action(s) that mend the situation at hand. The insight and/or corrective actions are passed to the *interested party* to get notified and take appropriate and crucial actions. The *interested party* component represents entities interested in the monitored element's health conditions which includes medical personnel and caregivers. These entities are provided with information of the current health conditions of the monitored element in distinct ways through the *presentation manager* component. A registration mechanism is provided by the presentation manager for the interested parties to register their interest in observing and tracing up the monitored element status. The so far presented concepts are then mapped onto software components at the design phase of the HMS process model. Also, at this phase, the relationships between these components are established which often involve a set of design patterns being applied. Figure 1 depicts the metamodel of the healthcare monitoring system.

#### C. *Functional requirements of proposed Framework*

This section is dedicated to introduce the functional requirements that represent the fundamental functions and services that must be provided by the Healthcare Monitoring Systems. These functions and services are listed below:
- *Register monitored element:* This function enables the HMS admin to register the monitored person on the system and give it a unique identifier.
- *Specify medical history for monitored element:* This function provides the HMS admin with the appropriate interface to specify and define the medical history of the monitored person. Such information represents the initial knowledge base of the monitored person.

- *Register device or sensor:* This function is used by the HMS admin to register the physical devices or sensors and assign to them a unique identifier.
- *Associate a monitored element with physical devices:* This is used by the HMS admin to associate each physical device and sensor that involves in collecting the data of the monitored person with the person of interest to the HMS.
- *Register interest with monitored element:* This function provides the interest party with the appropriate interface to register their interest in receiving notifications of the health conditions and medical current status of the monitored person.
- *Set context for monitored element:* This function is used to construct the current context of the monitored person. The context is composed of physiological data (e.g. temperature, blood pressure, etc) of interest to the healthcare monitoring system in question as well as the environmental data (e.g. room temperature) and activity data (e.g. moving, laying in bed, time of activity, location of activity, etc). All of these data are received from the three types of sensors depicted in figure 2.
- *Construct the context history:* This function enables the HMS to construct the *context history* which traces up the previous context data, which enables the HMS to be based on a series of observations and not only on the current situation presented by the current context data.
- *Create reasoning engine*: This function enables the HMS to create the reasoning engine which consists of machine learning algorithms that learn from the data defined in the *context history* and predict and produce possible *insights*.
- *Construct rule engine*: This function offers to the HMS admin the appropriate interface to create a set of rules that guides the actions performed as a response to any potential insight that represents the current status or the predicted or likelihood of the monitored person's health condition.
- *Create the presentation manager:* This function offers to the HMS admin the appropriate interface to create the presentation manager that is responsible for providing different views of the current status of the monitored person's health condition to different interested parties.

D. *Software components of proposed framework*

This section is dedicated to introduce the fundamental software components that constitute the proposed framework. Such components are derived from the functional requirements specified in the previous section.

A set of UML class diagrams is employed here to define each of these components as well as any relationships between them. The following are a description of these components:

- **MonitoredElement component:** This is the central component of the proposed framework on which the whole healthcare monitoring system is based. This is defined as a super class from which a set of subclasses inherits. The subclasses here include **Patient**, **Disabled** and **Elderly**. This class defines the common characteristics and behavior of its subclasses (personal information for instance)
- **MedicalHistory component:** This component is composed of a set of properties, in addition to a set of operations, to define the medical history of each instance of the **MonitoredElement component**. This component is part of the definition of the **MonitoredElement** component and thus is linked with it through the aggregation relationship. Each **MonitoredElement instance *has* one instance of MedicalHistory component.**
- **Sensor component**: This component is used to represent the common functionality of the sensor component. Specific sensors implement the sensor component functionalities differently, thus each sensor type should implement the interface of the sensor component by providing different behaviors in the *sense* operation. Also, the sensor component operates in two modes: time and event based modes.
- **MonitoredElementRegister component:** This component is used to register monitored elements (**Patient**, **Disabled** and **Elderly)** on the HMS. This class contains an important operation called r**egister** which takes as input a variable of **MonitoredElement** type. It also contains another operation called **unregister** to end the registration of monitored elements on the HMS.
- **SensorRegister component**: This component is used to register involved sensors on the HMS. This class contains an important operation called r**egister** which takes as input a variable of **Sensor** type. It has also another operation called **unregister** to end the registration of sensors on the HMS.
- **Association component**: This component contains the required operations to associate each sensor involving in the data collection process with the monitored element.

Such a component has an operation called `associate` which requires two variables, as input, of the `Sensor` and `MonitoredElement` types. Each instance of MonitoredElement type is associated with zero or more instances of `Sensor` type. A data structure of a map type (a key-value form) is used to establish this association where the *key* part is used to store one instance of `MonitoredElement` type and the *value* part is used to store a set of instances of Sensor type. Thus, the value part here should use a data structure that allows the storage of a collection of values (e.g. lists, sets).

- **`InterestedParty component:`** This component defines the common characteristics and behaviors of the different interested parties specified in the HMS. This is defined as a super class from which a set of subclasses inherits. The subclasses here include `CareGiver` and `MedicalPersonnel`.
- **`InterestRegistration component`**: This component is used by the HMS to allow different interested parties to register their interest in receiving notifications of the health conditions and current medical status of the `MonitoredElement`. The relationship between them can be established using the *observer* design pattern where the current medical status of `MonitoredElement` plays the role of the *subject* while each interested party plays the *observer* role.
- **`Context component:`** This component is used to define the context aspect of the `MonitoredElement`. Such a context is composed of three types of data: physiological data, environmental and activity data.
- **`ContextManager component`**: This component is in charge of managing the context aspect of the `MonitoredElement`. Operations of this component include *createContext, addToHistory, deleteHistory, etc.*
- **`ReasoningEngine component`:** This component is used to define the reasoning capability of the HMS that is needed to reason about the context history of the `MonitoredElement` for any likelihood of the occurrences of disease or any medical problems. This component contains an operation called *reason* that runs an appropriate machine learning algorithm that discovers possible *insights*. The functionality of this component is based mainly on two variables of `MedicalHistory and` ContextHistory types.
- **`PresentationManager component`**: This component is responsible for producing different views of the monitored element's medical status to different interested parties. Thus, the Model View Controller (MVC) design pattern can be applied here. The model here represents the data related to the current health conditions and status of the `MonitoredElement.` Any significant change should be reported to each `InterestedParty` using different views. The relationship between the model and view is established via the *observer* design pattern where the *model* plays the role of the *subject* while each *view* plays the *observer* role. The controller role here is played by the `ReasoningEngine component` which is responsible for changing the `MonitoredElement` status by discovering the current status or predicting an upcoming one.

E. *Software Architecture for Proposed Framework*

This section is dedicated to introduce the software architecture employed in this paper to design the Healthcare Monitoring Systems in the IoT platform. Abstractly speaking, the Healthcare Monitoring Systems consist of two fundamental entities: 1) the managed system which is represented here by the monitored element (patient, elderly and disabled) under observation and 2) the managing system whose main task is to take care of the patient or elderly under observation while in hospital or at home. The managing system is modelled in our previous work [9] using the control loop idea and following the IBM architecture blueprint. Such a control loop consists of four main activities namely the monitoring, analysis, planning and execution. These activities usually accomplish their tasks by using a knowledge base. What follows is a discussion of the layout and arrangement of the components responsible for the control loop activities over the Fog and cloud platforms. The following requirements have driven the proposed architecture in this paper:

- The management and control of the functioning of the IoT based Healthcare Monitoring System must be carried out as close to real time as possible.
- The support for the division of the local control loop into a number of smaller control loops where each one is tasked with managing and controlling a specific area in the same application to cover a wider area.
- The provision of powerful computing, storage and analysis capabilities in order to meet the

- requirements imposed by large scale and complex IoT applications.
- The support of the coordination between local control loops to regulate the functioning of the managed system in a decentralized mode.
- The delegation of some activities of the control loop to one or more local control loops and the regulation of the managed system in a centralized mode.

To meet the above identified requirements, we have suggested deploying a local control loop at the Fog platform in close proximity to the device or physical layer. This kind of local management is offered here as MAPEaaService. A slightly different service is also provided at the cloud platform to cater for the highly scaled and big data generating applications. This service contains only, in addition to the knowledge base, the analysis and planning activities and thus it is referred to as APaaService. Thus, the monitoring and execution activities are performed locally here. In this architecture, two modes of control are offered: centralized and decentralized. In the centralized mode, a central control loop is deployed either on the Fog or cloud platform depending on the scale of the application being designed and developed. Such a control loop regulates the functioning of the different control loops that reside at the same level at the Fog platform. The relationship between the central and local control loops is established using the master-slave model. The local control loop is responsible for monitoring and regulating parameters of high interest to a sub system of the whole system while the central control loop is in charge of controlling parameters of high interest to the whole system. In the Healthcare Monitoring Systems, the control loop of a sub system could be interested in monitoring the health condition of one patient by measuring and analyzing some specific parameters such as the temperature, blood pressure, blood sugar, etc. In contrast, the control loop of a whole system may have interest in finding out what the percentage of the affected patients of a specific disease within one particular area. Figure 2 shows a possible arrangement of a control loop with a centralized mode. Contrary to the centralized mode, the decentralized mode involves deploying a set of control loops of the same level where these loops coordinate to accomplish the four activities (monitoring, analysis, planning and execution) in the absence of a central control loop. Figure 3 depicts a the organisation of a control loop that adopts the decentralized mode.

## IV. RELATED WORK

A great deal of research papers and studies have been published in the area of Healthcare Monitoring Systems. This section is dedicated to the so far work carried out by researchers for engineering generic context aware frameworks for Healthcare Monitoring Systems in the IoT platform. Mshali [10], in his PhD thesis conducted a comprehensive study regarding the subject of this paper. The findings of his research revealed that the available of a full set of contextual information is crucial for providing patients and elderly people with required services and assistance in a proactive manner. He identified the important components required for designing and building context aware Healthcare Monitoring Systems. Chao et al [20] have proposed a monitoring system for patients of heart disease deployed on the IoT platform. Their system architecture is composed of the three typical layers: the sensing, network and application. Their system targets one specific kind of disease and does not take into consideration the issue of context awareness when designing such a system. Perera et al [21] have surveyed the important role that is played by context aware systems in supporting and boosting the IoT paradigm and its services and applications. Also, Alirezaie et al [22] have proposed a framework for long term monitoring of users of special needs using the IoT infrastructure. Such a framework exploits the semantic web idea to provide information with shared meaning across heterogeneous IoT devices, end users and healthcare professionals. It focuses on integrating measurements collected from different sources by using ontology to enable semantic interpretation of events and context awareness. In [23], the authors emphasized the importance of context awareness to the IoT paradigm in general and to healthcare monitoring systems in particular. They proposed an architecture that connects the different layers involved in the IoT platform to realize and deliver context aware healthcare monitoring systems. Another related work is introduced in [24] in which the authors developed an IoT based architectural framework with context awareness for hospital management systems. However, the main reason here for employing the context aware systems is to handle and manage the big data produced by the deployed sensors responsible for collecting the related data of monitored patients. Cisco [25] also has contributed to this area by proposing a context aware solution for healthcare systems. Such a solution enables hospitals to integrate contextual information such as location, temperature and presence information into the clinical workflow to increase staff efficiency, streamline inventory management and improve patient care.

To the best of our knowledge, none of the approaches proposed so far has studied the subject of developing Healthcare Monitoring Systems in the IoT platform using the control loops from the area of autonomic systems. Also, a few have recognised the importance of augmenting the management of such systems with the context awareness capability. Moreover, the arrangement of the management elements (control loop components) is often neglected in most studies.

## V. CONCLUSION

This paper has explored the vital role played by context aware systems in supporting IoT application in order to provide the right services at the right time. It has also investigated the importance of AI to the realization of resilient IoT applications and services. In particular, this paper tackled Healthcare Monitoring Systems and their provisioning in the IoT platform using ideas from AI field and in a context aware manner. We also highlighted the potential benefits that IoT can bring to healthcare systems as well as the challenges and risks of applying the IoT paradigm. More importantly, this paper introduced and proposed a generic framework for the design and development of context aware healthcare monitoring systems in the IoT platform. In such a framework, the fundamental components of the healthcare monitoring systems are identified and modelled as well as the relationship between them. This framework is based on a distributed layered architecture where the different components spread over the sensing layer, fog platform and the cloud platform

## VI. FUTURE WORK

For future work, the following issues need to be addressed:
- A detailed case study is needed to evaluate and illustrate the feasibility of the proposed framework.
- A more detailed design for the context element definition regarding the Healthcare Monitoring Systems.
- The investigation of different scenarios that illustrate the importance of choosing one control mode over the other (centralized or decentralized).
- The definition of a more detailed set of design patterns that helps system designers in designing and architecting Healthcare Monitoring Systems.

## ACKNOWLEDGMENT

The author greatly appreciate the precious support and encouragement received from the computer science department- Faculty of Science of Zintan University throughout the work on this paper.

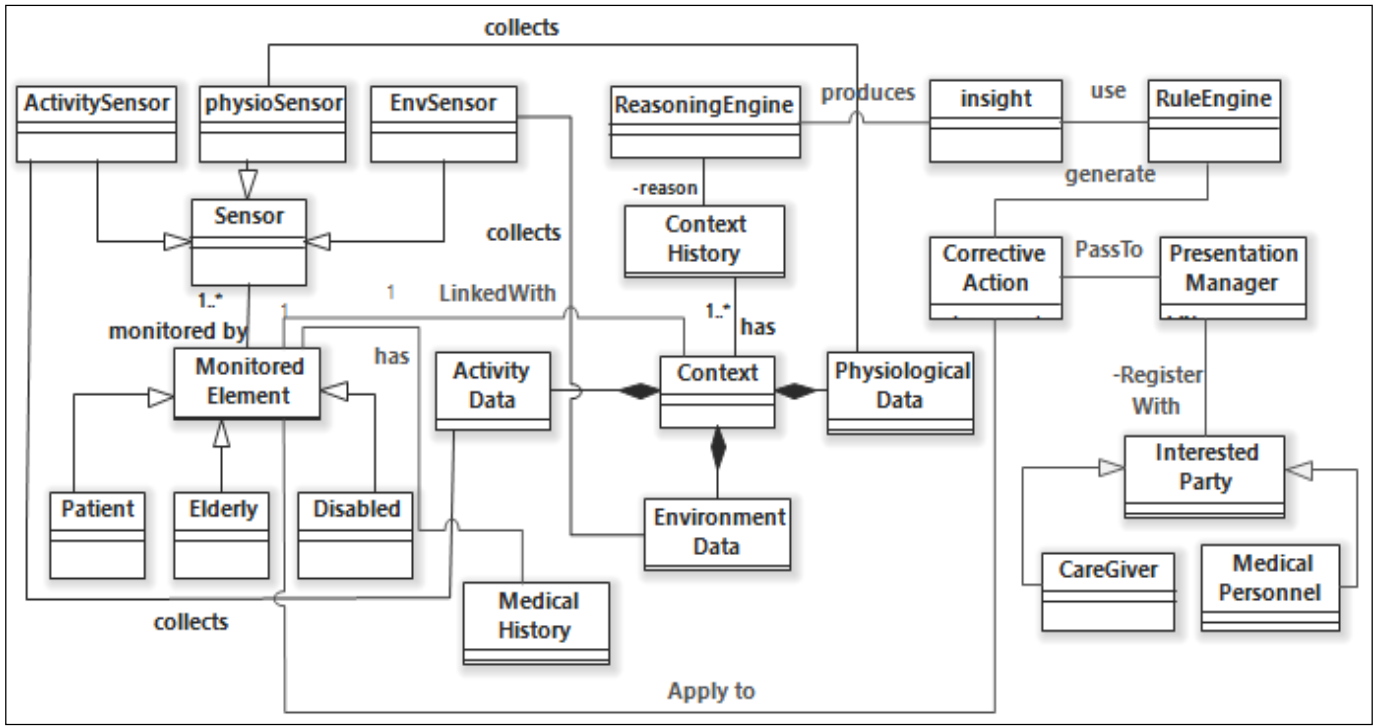

**Figure 1. A metamodel of proposed framework for Healthcare Monitoring Systems.**

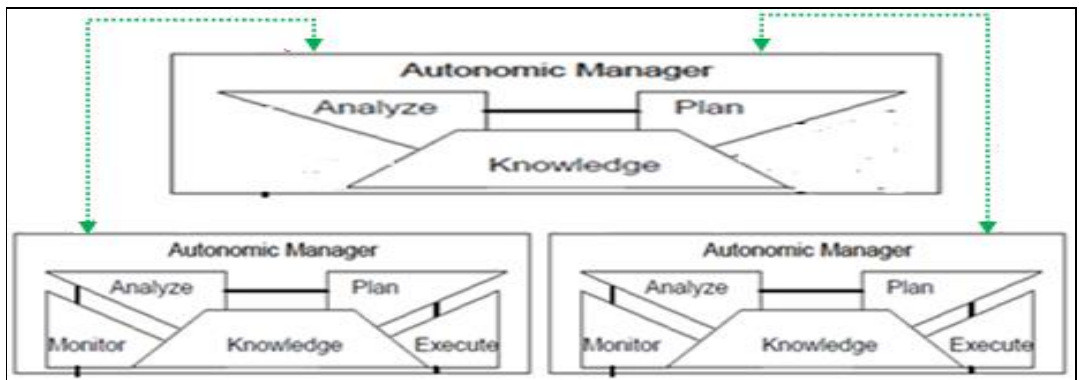

**Figure 2. A Centralized mode of control loop.**

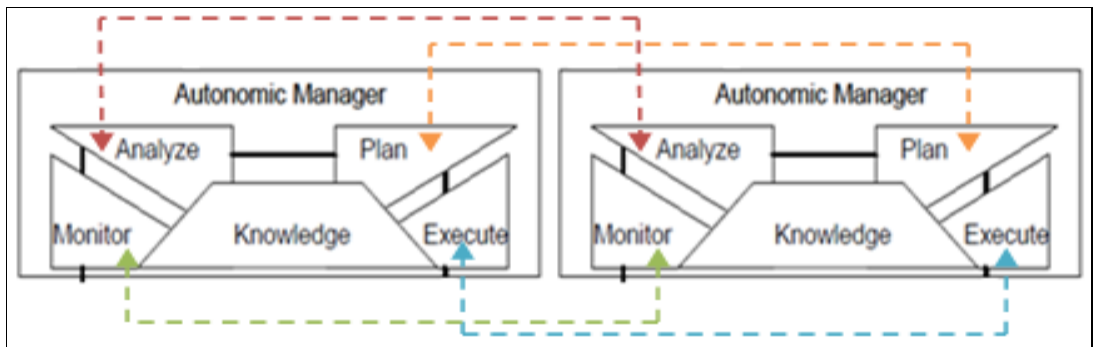

**Figure 3. A Decentralized mode of control loop**